\documentclass[12pt,a4paper]{article}
\usepackage[latin1] {inputenc}
\usepackage[T1]{fontenc}
\usepackage{amsmath}
\usepackage{amsfonts}
\usepackage{amssymb}
\usepackage[english]{babel}
\usepackage[dvips]{graphics,color}
\usepackage{relsize}
\usepackage{graphicx,psfrag,amsmath,amssymb,amsfonts,bbm,latexsym,color,dcolumn,
epsf}

\begin{document}
\title{Casimir effect with nonlocal boundary interactions} 
\author{C.~D.~Fosco and
E.~Losada\\
{\normalsize\it Centro At\'omico Bariloche and Instituto Balseiro}\\
{\normalsize\it Comisi\'on Nacional de Energ\'{\i}a At\'omica}\\
{\normalsize\it R8402AGP S. C. Bariloche, Argentina.} }
\maketitle
\begin{abstract}
We derive a general expression for the Casimir energy corresponding to two
flat parallel mirrors in $d+1$ dimensions, described by nonlocal
interaction potentials.  For a real scalar field, the interaction with the
mirrors is implemented by a term which is a quadratic form in the field,
with a nonlocal kernel.  The resulting expression for the energy is a
function  of the parameters that define the nonlocal kernel.  We show that
the general expression has the correct limit in the zero width case, and
also present the exact solution for a particular case.
\end{abstract}
\maketitle
The increasing interest in the Casimir effect~\cite{rev} is a natural
consequence of the availability of new precision experiments, which pose an
important pressure to continuously refine and improve the existing
calculations.  This evolution manifests itself at different levels; one of
them amounts to coping with situations where the geometry of the mirrors is
more complex, albeit with an idealized description  of their material
properties, i.e., they are regarded as mathematical surfaces occupied by
perfect conductors.
  Another level, which has recently received much attention, is the use of
a more accurate description of the mirrors, including corrections that
represent their departure from exactly conducting surfaces: rugosity,
finite temperature and conductivity, as well as finite width.  The latter
is usually dealt with by the introduction of a `space dependent mass term'
whereby the scalar field becomes very massive at the locii of the
mirrors~\cite{Graham:2003ib}.  For the case of a real scalar field
$\varphi$ in $d+1$ spacetime dimensions, and a single flat mirror centered
at $x_d=0$, the local Euclidean action, $S_{local}$, for this kind of term
is:  
\begin{equation}\label{eq:defsl}
S_{local}(\varphi) \;=\; \frac{1}{2} \, \int d^dx_\parallel \int dx_d \,
V_\epsilon (x_d) [\varphi(x_\parallel,x_d)]^2\;,
\end{equation}
where $x_\parallel$ denotes the time ($x_0$) as well as the $d-1$ spatial
coordinates parallel to the mirror (which we shall denote by
$\mathbf{x_\parallel}$).  The local potential $V_\epsilon(x_d)$ is a
positive function concentrated around $0$, with a width of size $\epsilon$.
The perfect mirror case is approached  when that size tends to zero and its
strength becomes infinite; namely, $V_\epsilon(x_d) \to  g \, \delta(x_d)$.
$g \to \infty$. This limit is a delicate step, since it usually introduces
divergences~\cite{Jaffe:2003ji} that may be difficult to deal with in a
setup that used idealized boundary conditions only. 

Local interaction terms have been extended to include a non trivial
dependence on the parallel coordinates~\cite{Saharian:2006zm,flm08}.
Translation invariance along them, necessarily implies a spatial
nonlocality:
\begin{equation}\label{eq:defsinl}
S_{local}'(\varphi) \;=\; \frac{1}{2} \, \int d^dx_\parallel \,\int
d^dx'_\parallel \int dx_d \; \varphi(x_\parallel,x_d) V_\epsilon(x_\parallel-x'_\parallel; x_d) 
\varphi(x'_\parallel,x_d) \;,
\end{equation}
but a Fourier transformation in $x_\parallel$, $x'_\parallel$ yields a
local expression in the mixed momentum ($k_\parallel$) and coordinate
($x_d$) representation: 
\begin{equation}\label{eq:defsinl1}
S_{local}'(\varphi) \;=\; \frac{1}{2} \, 
\int \frac{d^dk_\parallel}{(2\pi)^d} \int dx_d \; \widetilde{\varphi}^*(k_\parallel, x_d)\, 
\widetilde{V}_\epsilon(k_\parallel; x_d) \,
\widetilde{\varphi}(k_\parallel,x_d) \;,
\end{equation}
where the tildes denote the Fourier transformed of the corresponding
object.

Note that the resulting interaction term, $S_{local}'(\varphi)$, is still
assumed to be local in $x_d$, and  $\widetilde{V}_\epsilon(k_\parallel;
x_d)$ is concentrated around $x_d=0$, on a region of size $\sim \epsilon$.
However, except for the case of a zero-width mirror, a potential which is
local in $x_d$ can only be an approximate description of the interaction
with a real material. Indeed, as explained
in~\cite{EsquivelSirvent:2006iw}, one should in general use interactions
that also include `spatial dispersion' in $x_d$, i.e., nonlocality in the
normal coordinates. As also shown in ~\cite{EsquivelSirvent:2006iw}, in
spite of the nonlocality of the interaction, one may nevertheless use  a
Lifshitz formula~\cite{lifshitz} for the Casimir energy, since that formula
depends on the reflection coefficients at the media surfaces, and they may
be defined also for nonlocal media, under quite general assumptions. Interesting 
results have also been obtained using the boundary state formalism~\cite{bajnok}.

 We note in pass that the relevance of nonlocal media to Casimir-like
effects has also been appreciated in other contexts, like String
Theory~\cite{Elizalde:1995bb}. 

In spite of the validity and usefulness of Lifshitz formula for nonlocal
media, we believe that it would be important to have an alternative
expression for the Casimir energy, where its dependence on the details
defining the nonlocal media were more explicit.

One might also be interested in situations where the material is not
exactly confined to the region between two surfaces, but rather is
concentrated on a region, with a non zero (albeit rapidly vanishing)
density outside of that region. This is a situation where the use of
reflection coefficients, albeit still possible, becomes nevertheless
problematic.

To confront those problems, we first define the setup: for a mirror
centered at an arbitrary position $x_d=b$, we shall use an interaction term
$S_I^{(b)}$ given by: 
\begin{eqnarray}\label{eq:defsi}
S_I^{(b)}(\varphi) &=& \frac{1}{2} \, \int 
\frac{d^dk_\parallel}{(2\pi)^d} 
\int dx_d \; \int d{x'}_d  \nonumber\\
&\times&\widetilde{\varphi}^*(k_\parallel, x_d)\, 
\widetilde{V}_\epsilon(k_\parallel; x_d -b , {x'}_d - b) \,
\widetilde{\varphi}(k_\parallel,{x'}_d) \;.
\end{eqnarray}
The kernel $\widetilde{V}_\epsilon(k_\parallel; x_d, {x'}_d)$
is not invariant under translations in the normal direction 
($x_d \to x_d + h$, $x_d' \to x_d' + h$); besides, we still assume it to be
concentrated around $x_d=0$ and $x_d'=0$. 
This concentration may be used to write a convenient expansion for the
kernel, based on the introduction of $\big[ \psi_n^{(\epsilon)}(x_d)
\big]_n$, an orthonormal basis of functions of the normal coordinate,
obeying the boundary conditions that follow from the microscopic
model~\footnote{For example, they could satisfy Dirichlet boundary conditions at $x_d = \pm
\epsilon/2$. But they are not necessarily of compact support; they could
for example be exponentially decaying functions with a typical dispersion 
$\sim \epsilon$}.

 Then, without any loss of generality, the nonlocal kernel will be expanded as follows:
\begin{equation}\label{eq:quad1}
\widetilde{V}_\epsilon(k_\parallel; x_d, {x'}_d)\,=\,
\sum_{m,n} \, C_{mn}(k_\parallel,\epsilon) \, \psi_m^{(\epsilon)}(x_d)
\;\psi_n^{(\epsilon)*}({x'}_d) \;,
\end{equation} 
where $C_{mn}(k_\parallel,\epsilon)  = C_{nm}^*(k_\parallel,\epsilon)$, from the reality of the action. 

Being a nonlocal term, one has to rephrase the properties it should have to
behave as a (generalized) mass term, in the sense that it favours the
vanishing of the field around the region where it is different from zero.
It is clear then that the quadratic form (\ref{eq:quad1}) has to be
definite positive (in the space generated by the basis); this amounts to a
non-trivial relation for the $C_{mn}$ matrix.

Let us illustrate the previous construction with two examples. Firstly, we
consider a model where $S_I^{(b)}$ emerges from the linear coupling of
$\varphi$ to a microscopic real scalar field $\xi(x)$, which is confined to
the region $|x_d - b | \leq \epsilon/2$ ($x_\parallel$: arbitrary),
satisfying Dirichlet boundary conditions at $x_d = b \pm \epsilon/2$.  It
is sufficient to deal with $b=0$, since the general case is obtained by a
translation of the kernel.  Following a generalization of the approach
of~\cite{Fosco:2008td}, we see that, in the functional formalism,
$S_I^{(0)}$ may be written as follows:
\begin{equation}\label{eq:deseff}
e^{-S_I^{(0)}(\varphi)} \;=\; \frac{\int {\mathcal D}\xi \, e^{-S_m(\xi)
\,+\, i g \int d^{d}x_\parallel \int_{-\epsilon/2}^{+\epsilon/2} dx_d \, 
\xi(x_\parallel,x_d) \, \varphi(x_\parallel,x_d)}}{\int {\mathcal D}\xi \,
e^{-S_m(\xi)}} \;,
\end{equation}
where $g$ is a coupling constant and $S_m$ is the action for the
microscopic field. The matter field $\xi$ may have a
self-interaction, controlled by an independent coupling constant (implicit
in $S_m$).

To proceed, we denote by $W(J)$ the generating functional of connected
correlation functions of $\xi$, related to ${\mathcal Z}(J)$, the one for
the full correlation functions:
\begin{equation}
{\mathcal Z}(J)\,=\, \int {\mathcal D}\xi \, e^{-S_m(\xi) \,+\,\int
d^dx_\parallel \int_{-\epsilon/2}^{+\epsilon/2} dx_d \, 
J(x_\parallel,x_d) \xi(x_\parallel,x_d)} \;,
\end{equation}
by $W = \ln {\mathcal Z}$. The current $J$ is confined to the same region
as $\xi$, but it has free boundary conditions.
We then have that $S_I(\varphi) = - W[ i \,g\, \varphi(x)]$.
On the other hand, since only the quadratic part in $\varphi$ will be
retained~\footnote{The media are assumed to be linear.},
\begin{eqnarray}
S_I^{(0)}(\varphi) &=& -  W\big[i \,g\, \varphi(x)\big] \nonumber\\
&\simeq& \frac{1}{2} g^2 \, \int d^{d+1}x \int d^{d+1}x' \,
\varphi(x) W^{(2)}(x,x') \varphi(x') \;,
\end{eqnarray}
where $W^{(2)}$ is the connected $2$-point function.

Now, assuming that $S_m$ is translation invariant along $x_\parallel$, we
immediately identify the nonlocal kernel:
\begin{equation}
\widetilde{V}_\epsilon (k_\parallel;x_d,x_d') \;=\; g^2 \, {\widetilde
W}^{(2)}(k_\parallel;x_d,x_d') \;. 
\end{equation}
Besides, since $\xi$ satisfies Dirichlet boundary conditions, we have:
\begin{equation}
{\widetilde W}^{(2)}(k_\parallel; \pm\epsilon /2,x_d') \;=\; 
{\widetilde W}^{(2)}(k_\parallel;x_d,\pm\epsilon /2) \;=\; 0 \;.
\end{equation}
Then we have the expansion:
\begin{equation}\label{eq:dirichlet}
{\widetilde V}_\epsilon^{(0)}(k_\parallel;x_d,x_d') \;=\; 
\sum_{m,n} \, \psi_m^{(\epsilon)}(x_d) \, C_{mn}(k_\parallel,\epsilon) \,  \psi_m^{(\epsilon)*}(x_d) 
\end{equation}
where the orthonormal functions are given by:
\begin{equation}\label{eq:defpsi}
 \psi_n^{(\epsilon)}(x_d) \;=\; \sqrt{\frac{2}{\epsilon}} \,\times \, 
\left\{ 
\begin{array}{lcll}
\sin(\frac{n\pi x_d}{\epsilon}) & {\rm if}& \; n= 2 k, & (k = 1, 2,
\ldots)\\ 
\cos(\frac{n \pi x_d}{\epsilon}) & {\rm if}& \; n= 2 k + 1, & (k = 0, 1,
\ldots )\;.
\end{array}
\right.
\end{equation}

The precise form of $C_{mn}(k_\parallel,\epsilon)$ depends on the
action $S_m$. If it is a free action we have the {\em diagonal} expression:
\begin{equation}
 C_{mn}(k_\parallel,\epsilon) \;=\; 
\frac{g^2 \delta_{mn}}{ (\frac{n\pi}{\epsilon})^2 + k_\parallel^2 + \mu^2} \;,
\end{equation}
where $\mu$ is the mass of the microscopic field.

As an alternative example, we consider the case
of a charged field $\xi,\,\bar\xi$, coupled quadratically to the real field
$\varphi$. In this case, the analogue expression to (\ref{eq:deseff})
would be:
\begin{equation}\label{eq:seffcomp}
e^{-S_I^{(0)}(\varphi)} \;=\; \frac{\int {\mathcal D}\xi {\mathcal
D}\bar\xi \, e^{-S_m(\bar\xi,\xi)
\,+\, g \int d^{d}x_\parallel \int_{-\epsilon/2}^{+\epsilon/2} dx_d \, 
\bar\xi(x_\parallel,x_d)\varphi(x_\parallel,x_d) \xi(x_\parallel,x_d)}}{\int {\mathcal D}\xi {\mathcal
D}\bar\xi\, e^{-S_m(\bar\xi,\xi)}} \;,
\end{equation}
where, to simplify the treatment, we assume $S_m$ to be quadratic:
\begin{equation}
S_m(\bar\xi,\xi)\;=\; \int  d^{d}x_\parallel \int_{-\epsilon/2}^{+\epsilon/2} dx_d 
\big[ \partial\bar\xi \partial\xi \,+\, \mu^2 \bar\xi \xi \big]\;.
\end{equation}
The integral can be formally performed, but the result is a functional
determinant. Since we use a quadratic approximation for $S_I^{(0)}$, we
only need it up to that order in $\varphi$. The corresponding contribution
is just a $1$-loop diagram with two legs.  Translation invariance along the 
parallel coordinates suggest the use of a mixed Fourier representation:
\begin{equation}\label{eq:loop}
\tilde{V}_\epsilon(k_\parallel;x_d,x_d') \;=\; 
g^2 \, \int \frac{d^dp_\parallel}{(2\pi)^d} \, \tilde{G}(p_\parallel;x_d,x_d') 
\tilde{G}(p_\parallel + k_\parallel;x_d',x_d)
\end{equation}
where $\tilde{G}(p_\parallel;x_d,x_d')$ is the microscopic field propagator
in the mixed representation. 
Rather than evaluating the actual form of the nonlocal term for this model,
we just want to show that the boundary conditions for
the nonlocal kernel are determined from the ones we impose on the
microscopic field: assuming, for example, that this field
satisfies Dirichlet boundary conditions at the boundaries of the mirror,
from (\ref{eq:loop}) we derive for $\tilde{V}_\epsilon$ the
same kind of condition. Namely, the kernel vanishes when $x_d=\pm\epsilon /2$ or
$x_d'=\pm\epsilon /2$.  Thus, also in this case the model produces a
nonlocal kernel with the structure of (\ref{eq:dirichlet}) (the same
basis), with different coefficients $C_{mn}$. 

Before evaluating the energy, let us briefly examine the boundary
conditions that follow from the nonlocal term, in a concrete case. To that
end,  we consider the real-time version of the equations of motion for a
free massless scalar field coupled to a nonlocal potential centered at
$x_d=0$. Assuming, for the sake of simplicity, $d=1$, and $C_{mn} =
C_{mn}(\epsilon)$ (independent of $\omega$), the equation of motion
becomes:
\begin{equation}
\Box \varphi (x_0,x_1) \;=\;- \int dx_1' 
\, V_\epsilon(x_1,x_1')\,  \varphi(x_0,x_1') \;. 
\end{equation}
Fourier transforming in time,
\begin{equation}
(-\partial_1^2  - \omega^2) \tilde{\varphi}(\omega, x_1') 
\;=\;- \int dx_1' \, V_\epsilon(x_1,x_1')\,  \tilde{\varphi}(\omega,x_1') \;. 
\end{equation}
Then we multiply both sides of the equation above by
$\psi_m^{(\epsilon)*}(x_1)$ and integrate over $x_1$, to obtain:
 \begin{equation}
\langle \psi_m^{(\epsilon)} | (-\partial_1^2  -\omega^2)
|\tilde{\varphi}(\omega) \rangle 
\;=\;- \sum_n \, C_{mn}(\epsilon) \, 
\langle \psi_n^{(\epsilon)} |\tilde{\varphi}(\omega) \rangle \;, 
\end{equation}
where Dirac's  bracket notation denotes the scalar product of functions of
$x_d$.
Let us assume, for the sake of simplicity, that the proper basis
is (\ref{eq:defpsi}). Then $\langle \psi_m^{(\epsilon)} | (-\partial_1^2 )
|\tilde{\varphi}\rangle \,=\,  (\frac{m \pi}{\epsilon})^2 \, 
\langle \psi_m^{(\epsilon)} |\tilde{\varphi}\rangle$ and, as a consequence:
 \begin{equation}
\sum_n \, C_{mn}(\epsilon) \,\alpha_n
 \,=\, 
\big[\omega^2 - (\frac{m\pi}{\epsilon})^2 \big] \,\alpha_m ,
\end{equation}
where $\alpha_n \equiv \langle \psi_n^{(\epsilon)}
|\tilde{\varphi}(k_\parallel)\rangle$. 
Thus, 
 \begin{equation}
\sum_n \,C_{mn}(\epsilon) \,\alpha_m^* \, \alpha_n \,=\, 
\sum_m \big[\omega^2 - (\frac{m\pi}{\epsilon})^2 \big] \,|\alpha_m|^2 \;.
\end{equation}
This means that, to have a solution, the $\alpha_n$ coefficients
diagonalize $C$, and, since the quadratic form on the left hand side is positive, we
$\alpha_m$ vanishes whenever  $\omega^2 < (\frac{m\pi}{\epsilon})^2$. 

This means, in particular, that for $\omega^2 < (\frac{\pi}{\epsilon})^2$, all
the coefficients vanish: the field vanishes when $|x_1| < \frac{\epsilon}{2}$.
Of course, things are different if, for example: $(\frac{\pi}{\epsilon})^2 <
\omega^2 < (\frac{2\pi}{\epsilon})^2 $, then only the first coefficient may be
different from $0$. 

An interesting case is that of a $C_{mn}$ which is a finite matrix: one
that vanishes for $m> N$ of , say. An extreme case is $N=1$: there are then
only two regimes, depending on whether $\omega^2$ is bigger or smaller than
$(\frac{\pi}{a})^2$. In the former case, $\varphi$ in orthogonal to
$\psi_1^{(\epsilon)}$. This implies that {\em has at least one node\/} in
the $[-\frac{\epsilon}{2},\frac{\epsilon}{2}]$ interval. This is the
manifestation of a Dirichlet-like boundary condition in this context, which
of course will only hold true up to certain values of $\omega^2$. For bigger
values, the previous condition is relaxed and the mirror is transparent.  

Let us now evaluate the Casimir energy, discarding terms that are
independent of the distance between mirrors (and do not contribute to the
force).
For two mirrors, one at $x_d=0$ and the other at $x_d=a$, 
the total action $S$ is $S(\varphi) =S_0(\varphi)+S_I(\varphi)$ 
where $S_0 \equiv \frac{1}{2}\int d^{d+1}x \, \partial_\mu \varphi
\partial_\mu \varphi$, and $S_I(\varphi)=S_I^{(0)} + S_I^{(a)}$.

Since $S$ is a quadratic in the fields, it is immediate to find an
expression for the vacuum energy $E_0$, in terms of the determinant of the
corresponding kernel defining the quadratic form. Since we have translation
invariance along ${\mathbf x}_\parallel$, we use the energy per unit area, 
${\mathcal E}_0$, and take advantage of the Fourier transformation to obtain:
\begin{equation}
{\mathcal E}_0 \;=\; \frac{1}{2} \int \frac{d^dk_\parallel}{(2\pi)^d}\;
{\rm Tr}\ln \widetilde{\mathcal K}\;,
\end{equation}
where $\widetilde{\mathcal K}$ is an operator acting on functions of $x_d$, 
whose matrix elements are:
\begin{eqnarray}
\widetilde{\mathcal K}(x_d,x_d') &=& \widetilde{\mathcal K}_0(x_d,x_d') 
\,+\,  \widetilde{V}_\epsilon^{(0)}(k_\parallel; x_d , {x'}_d )
\nonumber\\
&+&  \widetilde{V}_\epsilon^{(a)}(k_\parallel; x_d , {x'}_d) \;. 
\end{eqnarray}
where ${\mathcal K}_0(x_d,x_d') \equiv (-\partial_d^2 + k_\parallel^2 )
\delta(x_d-x_d')$. 
The trace operation, denoted by `${\rm Tr}$' refers to the trace in the
space of functions depending on $x_d$.

The expression above contains three contributions which, to calculate the
Casimir force between the two mirrors, are irrelevant. One of them, ${\mathcal
E}_0^{vac}$, corresponds to the vacuum energy density in the absence of mirrors
\begin{equation}
{\mathcal E}_0^{vac} \,=\, 
\frac{1}{2} \int \frac{d^dk_\parallel}{(2\pi)^d}\;{\rm Tr}\ln
\widetilde{\mathcal K}_0 \;.
\end{equation}
The other two, denoted by ${\mathcal E}_0^{(0)}$ and ${\mathcal
E}_0^{(a)}$, are the mirrors' self-energies (and therefore we shall discard
them).  They have the form:
\begin{equation}\label{eq:selfener}
{\mathcal E}_0^{(b)} \;=\; 
\frac{1}{2} \int \frac{d^dk_\parallel}{(2\pi)^d}\;{\rm Tr} \ln 
\big( I + W^{(b)} \big) \;.
\end{equation}
where $b=0,\,a$, and $W^{(b)}$ denotes an operator, acting on the same
space as above, and defined by:
\begin{equation}
W^{(b)} \;=\; \widetilde{\mathcal K}_0^{-1} \, V^{(b)} \;.
\end{equation}
$V^{(0)}$, $ V^{(a)}$ have non trivial matrix
elements in smaller spaces, namely, the ones generated by the basis
functions sitting at each mirror. This fact can be used to show that the
trace operation above may be taken, for each term, using only the respective
basis at $x_d=0$ and $x_d=a$ (the trace operation over the complement
vanishes). 

Using matrix elements defined with the functions $\psi_n^{(\epsilon)}$ and
$\phi_n^{(\epsilon)}$, such that  
\mbox{$\phi_n^{(\epsilon)}(x_d) \equiv \psi_n^{(\epsilon)}(x_d - a)$}:
\begin{eqnarray}
W^{(0)}_{mn} &=& \langle \psi_m^{(\epsilon)}| \widetilde{\mathcal K}_0^{-1} \,
V^{(0)} | \psi_n^{(\epsilon)} \rangle \nonumber\\
W^{(a)}_{mn} &=& \langle \phi_m^{(\epsilon)}| \widetilde{\mathcal K}_0^{-1} \,
V^{(a)} | \phi_n^{(\epsilon)} \rangle \;.
\end{eqnarray}
$I$ denotes the identity operator, so that: $I_{mn}= \delta_{mn}$.
A simple shift of variables leads to ${\mathcal E}_0^{(0)} = {\mathcal
E}_0^{(a)}$, as it should be. 

After extracting the previous three contributions, we obtain a subtracted
energy density, $\tilde{\mathcal E}_0$, which by some straightforward 
algebra may be put in the form:
\begin{equation}
\tilde{\mathcal E}_0 \;=\; \frac{1}{2} \int \frac{d^dk_\parallel}{(2\pi)^d}\;
{\rm Tr}\ln \big( I - {\mathcal O} \big)\;,
\end{equation}
where ${\mathcal O}$ is an operator whose matrix elements may be
given in terms of the $\psi_n^{(\epsilon)}$ basis, as follows:
\begin{equation}\label{eq:defo}
{\mathcal O}_{mn} \;=\; \sum_{p,q,r} U_{m p}^{(0)} C_{pq} U_{q
r}^{(a)} C_{rn} \;,  
\end{equation}
where
\begin{eqnarray}
 U_{m n} &=& \langle \psi_m^{(\epsilon)}| \big[\widetilde{\mathcal K}_0 \,+\,
V_\epsilon^{(0)} \big]^{-1} | \phi_n^{(\epsilon)} \rangle \nonumber\\
 U_{m n}^{(a)} &=& \langle \phi_m^{(\epsilon)}| \big[\widetilde{\mathcal K}_0\,+\,
V_\epsilon^{(a)} \big]^{-1}| \psi_n^{(\epsilon)} 
\rangle \;=\;  U_{mn}^{(0)} \;\equiv\; U_{mn} \;.
\end{eqnarray}
Taking into account the previous relations,  
\begin{equation}
\tilde{\mathcal E}_0 \;=\; \frac{1}{2} \int \frac{d^dk_\parallel}{(2\pi)^d}\;
{\rm Tr}\ln \big( I - U \,C\, U \, C \big)\;,
\end{equation}
which is a nonlocal version, for flat, identical mirrors, of the equation derived in~\cite{Milton:2007wz} (see
also~\cite{Kenneth:2007jk}).

Note that $U_{m n}$ can be written more explicitly, by performing an
expansion in powers of $V^{(0)}_\epsilon$. Defining: 
\mbox{$\Delta_{mn}\equiv \langle \psi_m^{(\epsilon)}| \big[{\mathcal
K}_0]^{-1}| \psi_n^{(\epsilon)}\rangle $} and  
\mbox{$\Gamma_{mn}\equiv \langle \psi_m^{(\epsilon)}| \big[{\mathcal
K}_0]^{-1}| \phi_n^{(\epsilon)}\rangle $}, we see that:
\begin{eqnarray}
U_{m n}^{(0)}&=& \Gamma_{mn} - \Delta_{mp} C_{pq} \Gamma_{qn} \nonumber\\
&+& \Delta_{mp} C_{pq}\Delta_{qr} C_{rs} \Gamma_{sn} + \ldots
\end{eqnarray} 
(we used Einstein's summation convention). Then we
see that:
\begin{equation}
U \,=\, \big( I + \Delta C\big)^{-1} \Gamma \;.
\end{equation}
Let us check that the expressions above do yield the proper answer when the limit
corresponding to the case of perfect mirrors is taken. This is, in
the present formalism, tantamount to:
\begin{equation}\label{eq:vlocal}
\tilde{V}^{(0)}_\epsilon(k_\parallel;x_d,x_d') \;=\;  \lambda \; \delta(x_d) \;
\delta(x_d') \;,
\end{equation}
(sharp boundary conditions) and then $\lambda \to \infty$ (strong boundary
conditions). 
One immediately gets, from (\ref{eq:vlocal}), the matrix elements:
\begin{equation}\label{eq:vlocalmat}
C_{mn}\;=\;  \lambda \; \psi_m^{(\epsilon)}(0) \;
\psi_m^{(\epsilon)*}(0) \;. 
\end{equation}
Inserting this into (\ref{eq:defo}), we obtain:
\begin{equation}
\tilde{\mathcal E}_0\;=\; \frac{1}{2} \,
\int\;\frac{d^dk_\parallel}{(2\pi)^d}\,\ln \Big[1 \,-\,
\big(\frac{\lambda}{2 k_\parallel \,
+\, \lambda}\big)^2 \, e^{-2 k_\parallel a}\Big] \;,
\end{equation}
which in the strong limit, and for $d=3$ yields:
\begin{eqnarray}
\tilde{\mathcal E}_0&=& \frac{1}{2} \,
\int\;\frac{d^3k_\parallel}{(2\pi)^3}\,\ln \Big(1 \,-\, e^{-2 k_\parallel
a}\Big) \nonumber\\
&=& - \frac{\pi^2}{1440 \, a^3} \;,
\end{eqnarray}
which is the proper result.

Finally we consider a truly nonlocal example:
\begin{equation}
C_{mn}(k_\parallel,\epsilon) \;=\; \delta_{m0} \delta_{n0} \,
\lambda(k_\parallel) \;,
\end{equation}
where $\lambda(k_\parallel) > 0$.  It corresponds to a $1 \times 1$  matrix
$C$ in the space generated by the basis functions. In this case, we find that:
\begin{equation}
{\mathcal O}_{mn} \;=\; \big[\lambda(k_\parallel)\big]^2  \;U_{00} U_{m0} 
\delta_{n0}\;. 
\end{equation}
For this particular case, the ${\mathcal O}$ matrix has rank $1$, so only
one of its eigenvalues is different from $0$, and 
the trace of the log may be evaluated exactly. The Casimir energy becomes:
\begin{equation}
\tilde{\mathcal E}_0 \;=\; \frac{1}{2} \int \frac{d^dk_\parallel}{(2\pi)^d}\;
\ln \Big[ 1 -  \lambda^2(k_\parallel) \big(U_{00}\big)^2 \Big]\;.
\end{equation}
We may produce a more explicit expression for the matrix elements of $U$:
\begin{equation}
U_{00}^{(0)} = \big[ 1 + \lambda(k_\parallel) \,\Delta_{00} \big]^{-1} \,
\Gamma_{00} \;,
\end{equation}
where:
\begin{equation}
\Delta_{00} \,=\, \int \frac{dk_d}{2\pi} \, 
\frac{ \big|\langle \psi_0^{(\epsilon)}|k_d\rangle \big|^2 }{k_d^2 + k_\parallel^2} 
\end{equation}
($|k_d\rangle$ is the plane wave ket). 

The remaining object, $\Gamma_{00}$ is:
\begin{equation}
\Gamma_{00} \,=\, \int \frac{dk_d}{2\pi} \, e^{i k_d a} \,
\frac{ \big|\langle \psi_0^{(\epsilon)}|k_d\rangle \big|^2 }{k_d^2 +
k_\parallel^2} \;.
\end{equation}
Equipped with the previous expressions, we may calculate the Casimir energy
for a particular basis element. An interesting example is the choice of the
exponentially localized function
\begin{equation}
\psi_0^{(\epsilon)}=\frac{e^{-\frac{x^2}{2\,\epsilon^2}}}{\pi^{1/4}\,\epsilon^{1/2}}\;,
\end{equation}
since it allows one to evaluate the integrals above exactly. The result for
$\tilde{\mathcal E}_0$, in $d$ spatial dimensions, 
may be written in terms of an integral involving a function $G$
\begin{equation}
\tilde{\mathcal E}_0(a, \epsilon) \;=\;
\frac{1}{\Gamma\left(\frac{d}{2}\right) 2^d \pi^{d/2} a^d} \, \int
_0^\infty\;dp \, p^{d-1}\,
\ln\left\{ 1 -  \Big[ G\big( p; \frac{2 \epsilon}{a}\,,\,
a \lambda(\frac{p}{a}) \big)\Big] ^2\right\} \;,
\end{equation}
depending on dimensionless parameters. It is given explicily by:
\begin{equation}
G(p;x,l)\,=\,\frac{e^{-p}\, {\rm erfc}( x\, p - \frac{1}{2 x} ) + \,e^p\,{\rm erfc}( x\, p
+ \frac{1}{2 x})}{\frac{p\, e^{- x^2 p^2 } }{2\,\sqrt{\pi}\,l} + 2\,{\rm erfc}\left( x p\right) }\,,  
\end{equation}
where ${\rm erfc}$ is the complementary error function.

It is immediate to check that the Casimir energy for the perfect mirror
case is reproduced when $\frac{\epsilon}{a} \to 0$ and $\lambda \to
\infty$. On the other hand, when that limit is taken for a {\em finite\/}
$\lambda$, the result has the same
form as in the local case, but with a (finite) renormalization for
$\lambda$. Indeed, if $\lambda_{local}$ denotes the coupling constant in the local
$\delta$-potential case, the energies agree for $\lambda = \frac{\lambda_{local}}{8\sqrt{\pi}}$. 

For finite values of $\frac{\epsilon}{a}$, we have the interesting
phenomenon that the corrections are not analytical in that variable.
Indeed, one can see that the corrections to the zero-width case
are proportional to a factor $e^{- (\frac{a}{2\epsilon})^2}$. 

To make the comparison with the local case more explicit we present, in
Figure 1, the plots of  the Casimir energies corresponding to the local and
nonlocal cases, for $d=1$. The local potential is chosen so that it agrees
with the nonlocal one when $\epsilon \to 0$ ($\lambda_{local} = 8\sqrt{\pi}
\lambda$):
\begin{figure}
\begin{center}
\begin{picture}(0,0)%
\includegraphics{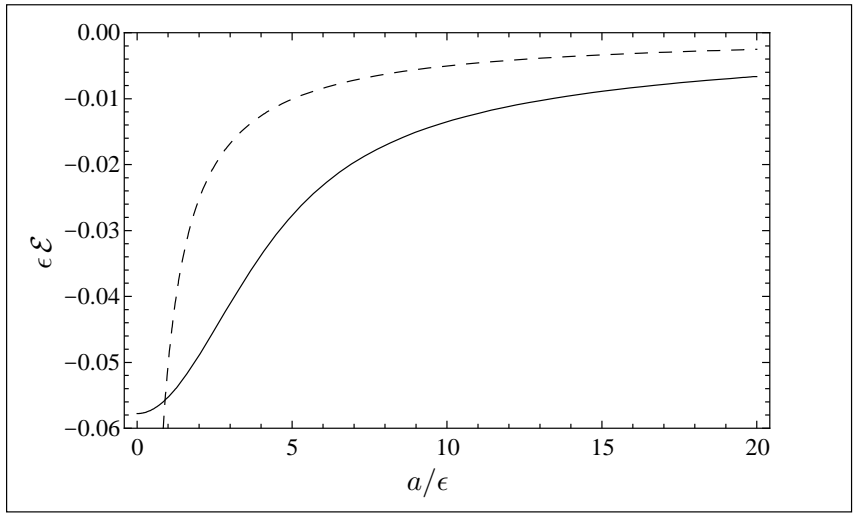}%
\end{picture}%
\setlength{\unitlength}{3947sp}%
\begingroup\makeatletter\ifx\SetFigFontNFSS\undefined%
\gdef\SetFigFontNFSS#1#2#3#4#5{%
  \reset@font\fontsize{#1}{#2pt}%
  \fontfamily{#3}\fontseries{#4}\fontshape{#5}%
  \selectfont}%
\fi\endgroup%
\begin{picture}(5274,3201)(2014,-6457)
\put(2302,-4926){\rotatebox{90.0}{\makebox(0,0)[lb]{\smash{{\SetFigFontNFSS{11}{13.2}{\rmdefault}{\mddefault}{\updefault}{\color[rgb]{0,0,0}$\epsilon \,\mathcal{E}$}%
}}}}}
\put(4513,-6307){\makebox(0,0)[lb]{\smash{{\SetFigFontNFSS{11}{13.2}{\rmdefault}{\mddefault}{\updefault}{\color[rgb]{0,0,0}$a/\epsilon$}%
}}}}
\end{picture}%
\end{center}
\caption{Casimir energies corresponding to the nonlocal (continuous line)
and local (dashed line) cases, as a function of $b \equiv
\frac{a}{\epsilon}$ for $\epsilon \lambda \equiv \frac{1}{8 \sqrt{\pi}}$.}
\end{figure}

A remarkable fact, that can be observed in the plot, is that the Casimir
energy for the nonlocal case becomes {\em finite\/} when the distance between the
mirrors tends to zero. This is a manifestation of the fact that the
nonlocality softens the UV behaviour of the system. Yet another consequence
of the same effect is that the integral over $k_\parallel$ for the energy of a single mirror, 
${\mathcal E}_0^{(b)}$ (see Eq. (\ref{eq:selfener})), has a better UV behaviour that its local
counterpart. In particular, for $d=1$, a simple calculation shows that it
becomes: 
\begin{equation}
{\mathcal E}_0^{(b)} \;=\; \frac{1}{2\pi}\int_{0}^{\infty} dk\;
\ln \left[ 1 + \, \frac{4 \lambda \sqrt{\pi}}{k}\;\;e^{ 4\,\epsilon ^2
k^2}\,{\rm erfc}\left(2 \epsilon \,k\right) \right] \;,
\end{equation}
which is convergent for any $\epsilon > 0$ (we recall that  its local counterpart is
logarithmically divergent~\cite{Jaffe:2003ji}). 

We conclude by noting that, as shown in the examples above, nonlocal
potentials can be used to impose  boundary conditions in finite size
mirrors, and they becomes automatically frequency dependent. 
Also, in spite of their seemingly complex structure, a general expression
for the energy may be derived, which contains some new interesting features:
the non-analytic behaviour of its small-width expansion and a softer UV
behaviour.

In spite of the above, the perfect mirror limit is still properly reproduced. Besides,
when the distance between mirrors is of the order of $\epsilon$, the
Casimir force vanishes, rather than becoming infinite, as it happens in the
local case. 

\section*{Acknowledgements}
We thank Prof.\ F.\ D. Mazzitelli for many useful comments and discussions. 
This work was partially supported by CONICET, ANPCyT and UNCuyo. 

\end{document}